\documentclass[journal]{IEEEtran}
\usepackage{amsmath,amsfonts}
\usepackage{algorithmic}
\usepackage{algorithm}
\usepackage{array}
\usepackage[caption=false,font=normalsize,labelfont=sf,textfont=sf]{subfig}
\usepackage{textcomp}
\usepackage{stfloats}
\usepackage{url}
\usepackage{verbatim}
\usepackage{graphicx}
\usepackage{cite}
\usepackage{color}
\usepackage{amssymb}
\usepackage{hyperref}
\begin{document}

\title{On the Performance Analysis of Pinching-Antenna-Enabled SWIPT Systems}
\author{Bingxin~Zhang,~\IEEEmembership{Member,~IEEE,}
	Han~Zhang,~\IEEEmembership{Student~Member,~IEEE,}
	Kun~Yang,~\IEEEmembership{Fellow,~IEEE,}
	~Yizhe~Zhao,~\IEEEmembership{Member,~IEEE,}
	and Kezhi~Wang,~\IEEEmembership{Senior~Member,~IEEE}
	\thanks{Bingxin Zhang, Han Zhang and Kun Yang  are with the State Key Laboratory of Novel Software Technology, Nanjing University, Nanjing 210008, China, and School of Intelligent Software and Engineering, Nanjing University (Suzhou Campus), Suzhou, 215163, China (email: bxzhang@nju.edu.cn; hanzhangl@smail.nju.edu.cn; kunyang@nju.edu.cn).}
	\thanks{Yizhe Zhao is  with the School of Information and Communication Engineering, University of Electronic Science and Technology of China, Chengdu 611731, China (e-mail: yzzhao@uestc.edu.cn).}
	\thanks{Kezhi Wang is with Department of Computer Science, Brunel University London, Uxbridge, Middlesex, UB8 3PH (email: kezhi.wang@brunel.ac.uk).}
}



\maketitle

\begin{abstract}
In this paper, we studies the performance of a novel simultaneous wireless information and power transfer (SWIPT) system enabled by a flexible pinching-antenna. To support flexible deployment and optimize energy-rate performance, we propose three practical pinching antenna placement-schemes: the edge deployment scheme (EDS), the center deployment scheme (CDS), and the diagonal deployment scheme (DDS). Moreover, a hybrid time-switching (TS) and power-splitting (PS) protocol is introduced, allowing dynamic adjustment between energy harvesting and information decoding. Under each deployment strategy and the transmission protocol, closed-form expressions for the average harvested energy and average achievable rate of a randomly located user equipment (UE) are derived based on the optimal positioning of the pinching-antenna. Numerical simulations confirm the accuracy of the theoretical analysis and illustrate the trade-off between rate and energy harvesting under different schemes.

\end{abstract}


\section{Introduction}
The rapid growth of wireless data traffic and the emergence of energy-constrained devices in 6G networks have necessitated the integration of simultaneous wireless information and power transfer (SWIPT) into communication systems \cite{HJYK}. However, existing SWIPT techniques face significant challenges when deployed in complex environments with severe path loss or non-line-of-sight (NLoS) conditions. To address these issues, several flexible antenna architectures have been proposed in recent years, including intelligent reflecting surfaces (IRS) \cite{WQZR}, fluid antennas \cite{WKK} , and mobile antennas \cite{ZLMW}. IRSs can construct virtual line-of-sight (LoS) links by intelligently reflecting signals. However, its performance is inherently limited by the double fading effect. Fluid antennas and movable antennas provide physical relocation capability; however, their movement range is typically constrained within a few wavelengths, making them ineffective against large-scale fading or user mobility over extended areas. 

To overcome these limitations, a novel concept termed the pinching-antenna has been proposed by NTT DOCOMO \cite{AFHY}, which enables dynamic channel configuration by physically placing a dielectric “clip” at arbitrary positions along a flexible waveguide. This promising architecture is expected to play a key role in future wireless systems by enhancing line-of-sight connectivity, mitigating large-scale fading, and enabling reconfigurable antenna deployment tailored to user mobility and energy demands \cite{LYWZ}.

Although the research on pinching-antennas is still in its early stages, several initial studies have been conducted to explore their fundamental properties and performance in wireless systems. The authors in \cite{DZSR} analytically investigates the performance of pinching-antenna systems under various configurations, demonstrating their potential to enhance LoS links, support NOMA, and approach the performance limits of MISO interference channels. In \cite{XJWJ}, the authors were the first to investigate the channel estimation problem in pinching-antenna systems and proposed two efficient deep learning-based channel estimation algorithms. In \cite{ZZYZ}, the authors further jointly optimized the users' power location coefficients and the positions of the pinching-antennas. The authors in \cite{TDTSA} proposed a unified analytical framework to characterize the outage probability and average rate performance of pinching-antenna systems.

However, no existing work has explored the potential benefits of employing pinching-antennas in SWIPT systems. To fill this gap, we consider a pinching-antenna-enabled SWIPT system in this paper. Specifically, the main contributions of our work are summarized as follows: (i) we introduce the pinching antenna into the SWIPT system, propose three waveguide deployment schemes, and consider a hybrid time switching (TS) and power splitting (PS) transmission protocol; (ii) For each deployment scheme, we derive novel closed-form expressions for the average harvested energy and average achievable rate under both linear energy harvesting model (LM) and non-linear energy harvesting model (NLM), offering analytical insights into the energy-rate trade-off; (iii) Extensive Monte-Carlo simulations validate the accuracy of the derived expressions and reveal how waveguide deployment schemes and transmission protocol parameters influence the trade-off between rate and energy performance.
\begin{figure}[!t]
	\centering
	\includegraphics[width=2.18 in]{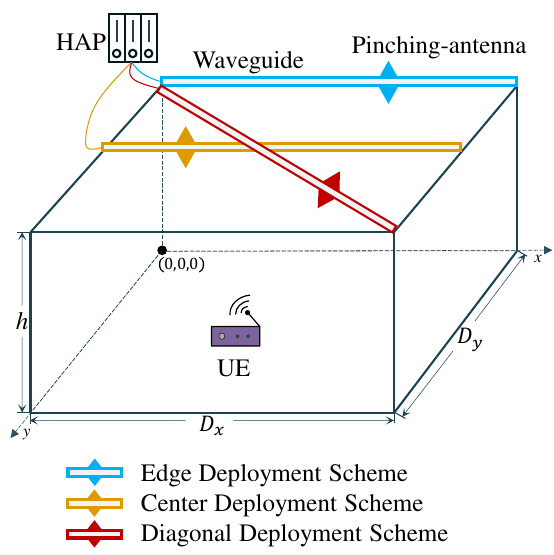}
	\caption{System model of the pinching-antenna-enabled SWIPT system.}
	\label{sysmod}
\end{figure}
\section{System Model}
As illustrated in Fig. \ref{sysmod}, we consider a SWIPT system in which a base station (BS) serves a single-antenna user equipment (UE). In particular, the UE is assumed to be randomly located within a rectangular region on the $x$–$y$ plane, with side lengths $D_x$ and $D_y$ along the $x$- and $y$-axes, respectively. The UE’s position is denoted by $\boldsymbol{\psi}_{\mathrm{u}} = (x_{\mathrm{u}}, y_{\mathrm{u}}, 0)$, where $x_{\mathrm{u}} \sim \mathcal{U}[0, D_x]$ and $y_{\mathrm{u}} \sim \mathcal{U}[0, D_y]$. To ensure robust information and energy transmission, the BS employs a pinching-antenna integrated with a dielectric waveguide, enabling dynamic adjustment of the antenna's position along the waveguide to optimize the wireless transmission link. Unlike conventional wireless architectures, the pinching-antenna system facilitates low-loss propagation of high-frequency signals within the waveguide and enables controlled radiation at arbitrary locations along its length via a ``pinching” mechanism. Specifically, the waveguide is deployed at a height $h$, and the positions of the BS and the pinching-antenna are given by $\boldsymbol{\psi}_{\mathrm{b}} = (0, 0, h)$ and $\boldsymbol{\psi}_{\mathrm{p}} = (x_{\mathrm{p}}, y_{\mathrm{p}}, h)$, respectively. It is worth noting that three waveguide deployment schemes are proposed, namely edge deployment scheme, center deployment scheme, and diagonal deployment scheme, as depicted in Fig. \ref{sysmod}. The detailed configurations of each scheme will be discussed in Section III.

Furthermore, a hybrid time-switching (TS) and power-splitting (PS) protocol is employed. In this scheme, the parameters $T$, $\alpha \in \left[ 0, 1 \right] $ and $\beta \in \left[  0, 1 \right]$ correspond to the total transmission period, TS factor and PS factor, respectively. To facilitate analysis, $T$ is normalized to 1.

Therefore, the received signal at the UE can be given by
\begin{equation}\label{formula_1}
	\begin{aligned}
		y = \frac{ \sqrt{\mu P_t}  e^{-j \frac{2 \pi}{\lambda} \left\| \boldsymbol{\psi}_\mathrm{p} - \boldsymbol{\psi}_\mathrm{u} \right\| }}{\left\| \boldsymbol{\psi}_\mathrm{p} - \boldsymbol{\psi}_\mathrm{u} \right\|} s + n,
	\end{aligned}
\end{equation}
where $\mu = \frac{c^2}{16 \pi^2 f_c^2}$, $c$ represents the speed of light, $f_c$ denotes the carrier frequency, $P_t$ represents the transmit power for UE's signal, $\lambda$ is the free-space wavelength, $s$ is the transmitted signal by the BS satisfying $\mathbb{E} \left[ |s|^2 \right] = 1$, $n \sim \mathcal{CN} (0, \sigma^2)$ denotes the additive white Gaussian noise with zero mean and variance $\sigma^2$, $\left\| \boldsymbol{\psi}_\mathrm{p} - \boldsymbol{\psi}_\mathrm{u} \right\|$ denotes the distance between the pinching antenna and the UE.

For energy harvesting, based on the transmission protocol and assuming a LM, the average harvested energy at the UE can be expressed as
\begin{equation}\label{formula_2}
	\begin{aligned}
		\bar{E}_{\mathrm{LM}} = \mathbb{E} \left[ \frac{\alpha \beta \eta P_t}{\left\| \boldsymbol{\psi}_\mathrm{p} - \boldsymbol{\psi}_\mathrm{u} \right\|^2} \right], 
	\end{aligned}
\end{equation}
where $\eta \in (0, 1)$ denotes the energy conversion efficiency coefficient of the UE's rectifying antenna.

However, in practice, the energy harvesting circuits in the UE typically exhibit non-linear characteristics due to components like diodes or capacitors. To more accurately describe the actual energy harvesting process, a logistic model is widely adopted. Under the NLM, the energy harvested at the UE can be redefined as
\begin{equation}\label{formula_3}
	\begin{aligned}
		{\Phi} (P_\mathrm{in}) = \left[ \frac{\varphi}{1 - \Omega} \left( \frac{1}{1 + e^{-a (P_\mathrm{in} - b)}} - \Omega \right) \right]^+, 
	\end{aligned}
\end{equation}
where $[x]^+ = \max (0, x)$, $\Omega = \frac{1}{1 + e^{a b}}$, $P_\mathrm{in}$ represents the incident power, the constants $\varphi$, $a$ and $b$ are determined by the characteristics of the energy harvesting circuit, such as diodes and resistances. 

Consequently, by adopting the NLM in (\ref{formula_3}), the average energy harvested at the UE can be written as
\begin{equation}\label{formula_4}
	\begin{aligned}
		\bar{E}_{\mathrm{NLM}} = \alpha \mathbb{E} \left[ \Phi(P_\mathrm{in}) \right], 
	\end{aligned}
\end{equation}
where $P_\mathrm{in} = \frac{\beta P_t}{\left\| \boldsymbol{\psi}_\mathrm{p} - \boldsymbol{\psi}_\mathrm{u} \right\|^2}$.

According to (\ref{formula_1}), the received signal-to-noise ratio (SNR) can be expressed as
\begin{equation}\label{formula_5}
	\begin{aligned}
		\gamma = \frac{\mu P_t \left| e^{-j \rho \left\| \boldsymbol{\psi}_\mathrm{p} - \boldsymbol{\psi}_\mathrm{u} \right\| }\right|^2 }{\left\| \boldsymbol{\psi}_\mathrm{p} - \boldsymbol{\psi}_\mathrm{u} \right\|^2 \sigma^2}.
	\end{aligned}
\end{equation}

With respect to information decoding, according to the hybrid transmission protocol, the average achievable rate at the UE can be defined as 
\begin{equation}\label{formula_6}
	\begin{aligned}
		\bar{R} &= \left[ \alpha (1 - \beta) + (1 - \alpha) \right] \mathbb{E} \left[ \log_2 \left(1 + \gamma \right) \right] \\
		&= (1 - \alpha \beta) \mathbb{E} \left[ \log_2 \left(1 + \frac{\mu \bar{\gamma}}{\left\| \boldsymbol{\psi}_\mathrm{p} - \boldsymbol{\psi}_\mathrm{u} \right\|^2 } \right) \right], 
	\end{aligned}
\end{equation}
where $\bar{\gamma} = \frac{P_t}{\sigma^2}$.

\section{Performance Analysis}
In the investigated pinching-antenna-enabled SWIPT system, we propose three waveguide deployment schemes:
\begin{itemize}
	\item \textbf{Edge Deployment Scheme (EDS):} The waveguide is deployed at height $h$, aligned parallel to the $x$-axis, with a fixed $y$-coordinate of 0. Accordingly, the coordinate of the pinching-antenna is denoted as $(x_{\mathrm{p},1}, 0, h)$, where $x_{\mathrm{p},1} \in \left[ 0, D_x \right]$.
	
	\item \textbf{Central Deployment Scheme (CDS):} The waveguide is deployed at height $h$, also parallel to the $x$-axis, but fixed at $y = \frac{D_y}{2}$. The corresponding pinching-antenna coordinate is given by $(x_{\mathrm{p},2}, \frac{D_y}{2}, h)$, where $x_{\mathrm{p},2} \in \left[ 0, D_x \right]$.
	
	\item \textbf{Diagonal Deployment Scheme (DDS):} The waveguide is deployed along the top diagonal direction at height $h$, and the pinching-antenna position is expressed as $(x_{\mathrm{p},3}, y_{\mathrm{p},3}, h)$, where $x_{\mathrm{p},3} \in \left[ 0, D_x \right]$, $y_{\mathrm{p},3} = \frac{D_y}{D_x} x_{\mathrm{p},3}$.
\end{itemize}

To evaluate the performance of the average harvested energy and average achievable rate, it is essential to first characterize the distribution of the distance between the pinching antenna and the UE under the three waveguide deployment schemes, denoted as \( L_{\nu} = \left\| \boldsymbol{\psi}_{\mathrm{p},\nu} - \boldsymbol{\psi}_\mathrm{u} \right\| \), where \( \nu \in \{1, 2, 3\} \). 

Leveraging the flexibility of the pinching-antenna, it is strategically positioned as close to the UE as possible to maximize the SWIPT performance by minimizing the impact of large-scale fading. Accordingly, under the EDS and CDS configurations, the optimal pinching-antenna positions are denoted by $\boldsymbol{\psi}_\mathrm{p,1}^* = (x_\mathrm{u}, 0, h)$ and $\boldsymbol{\psi}_\mathrm{p,2}^* (x_\mathrm{u}, \frac{D_y}{2}, h)$, respectively. For the DDS, the distance between the pinching-antenna and the UE is derived as
\begin{equation}\label{formula_7}
	\begin{aligned}
		L_3 &= \left\| \boldsymbol{\psi}_{\mathrm{p}, 3} - \boldsymbol{\psi}_\mathrm{u} \right\| = \left( x_{\mathrm{p},3} - x_{\mathrm{u}} \right)^2 + \left(\frac{D_y}{D_x} x_{\mathrm{p},3} - y_{\mathrm{u}} \right)^2 + h^2 \\
		&= (1 + k^2) x_{\mathrm{p},3}^2 - 2 (x_{\mathrm{u}} + k y_{\mathrm{u}}) x_{\mathrm{p},3} + (x_{\mathrm{u}}^2 + y_{\mathrm{u}}^2 + h^2), 
	\end{aligned}
\end{equation}
where $k = \frac{D_y}{D_x}$.

Then, by taking the derivative of $x_{\mathrm{p},3}$ in (\ref{formula_7}) and setting it to zero, i.e., $\frac{d L_3}{d x_{\mathrm{p},3}} = 0 $, we obtain
\begin{equation}\label{formula_8}
	\begin{aligned}
		x_{\mathrm{p}, 3}^* = \frac{x_{\mathrm{u}} + k y_{\mathrm{u}}}{1 + k^2} \quad \textrm{and} \quad y_{\mathrm{p}, 3}^* = \frac{k x_{\mathrm{u}} + k^2 y_{\mathrm{u}}}{1 + k^2}.
	\end{aligned}
\end{equation}
Accordingly, the optimal position of the pinching-antenna under the DDS is given by $L_3^* = (x_{\mathrm{p}, 3}^*, y_{\mathrm{p}, 3}^*, h)$.

Then, based on the optimal pinching-antenna configurations under the three deployment schemes, the CDF and PDF of the random variable $L_\nu^*$, with $\nu \in \left\lbrace 1, 2, 3 \right\rbrace$, will be presented in the following two lemmas.

\textbf{Lemma 1}. For the EDS and CDS, the CDF and PDF of the random variable $L_{\varpi}$ are given by
\begin{equation}\label{formula_9}
	{F_{{L_{\varpi}^*}}}(l) = \left\{ {\begin{array}{*{20}{l}}
			{0,} & {l < h^2}, \\
			{\frac{\varpi \sqrt{l - h^2}}{D_y},} & {h^2 \le l \le h^2 + \left(\frac{D_y}{\varpi}\right)^2}, \\
			{1,}&{l > h^2 + \left(\frac{D_y}{\varpi}\right)^2},
	\end{array}} \right.
\end{equation}
and
\begin{equation}\label{formula_10}
	{f_{L_{\varpi}^*}} \left( l \right) = \left\{ {\begin{array}{*{20}{c}}
			\frac{\varpi}{2 D_y \sqrt{l - h^2}}, & { h^2 \le l \le h^2 + \left(\frac{D_y}{\varpi}\right)^2 }, \\
			0, &{{\rm{otherwise}}},
	\end{array}} \right.
\end{equation}
respectively, $\varpi \in \{1, 2\}$.

\begin{IEEEproof}
We first analyze the cumulative distribution function (CDF) of the minimum distance $L_1^*$ between the pinching-antenna and the UE under the EDS. Therefore, one has
\begin{equation}\label{formula_11}
	L_1^* = y_{\mathrm{u},1}^2 + h^2.
\end{equation}

Then, given that $y_{\mathrm{u}} \sim \mathcal{U}[0, D_y]$ and the definition $F_{L_1^*} = \Pr \left\lbrace L_1^* \leq l \right\rbrace $, the CDF of $L_1^*$ can be readily obtained as shown in (\ref{formula_9}). Furthermore, taking the first-order derivative of (\ref{formula_9}) with respect to $l$ yields the PDF of $L_1^*$, as given in (\ref{formula_10}). Finally, the CDF and PDF of the random variable $L_2^*$ can be obtained using a method similar to that of $L_1^*$.
\end{IEEEproof}

\textbf{Lemma 2}. For the DDS, the CDF and PDF of $L_3$ can be derived as
\begin{equation}\label{formula_12}
	{F_{{L_3^*}}}(l) = \left\{ {\begin{array}{*{20}{l}}
			{0,} & {l < h^2}, \\
			{\frac{2 \Lambda \sqrt{l - h^2} - l + h^2}{\Lambda^2},} & {h^2 \le l \le h^2 + \Lambda^2}, \\
			{1,}&{l > h^2 + \Lambda^2}, 
	\end{array}} \right.
\end{equation}
and
\begin{equation}\label{formula_13}
	{f_{L_3^*}} \left( l \right) = \left\{ {\begin{array}{*{20}{c}}
			\frac{1}{\Lambda \sqrt{l - h^2}} - \frac{1}{\Lambda^2}, & { h^2 \le l \le h^2 + \Lambda^2 }, \\
			0, &{{\rm{otherwise}}},
	\end{array}} \right.
\end{equation}
respectively, where $\Lambda = \frac{D_x D_y}{\sqrt{D_x^2 + D_y^2}}$.

\begin{IEEEproof}  
Please refer to Appendix \ref{proofone}.
\end{IEEEproof}

\textbf{Lemma 3}. For the EDS and CDS, the average harvested energy at the UE under the LM can be expressed as
\begin{equation}\label{formula_14}
	\begin{aligned}
		\bar{E}_{\mathrm{LM}, \varpi} = \frac{\alpha \beta \eta \varpi P_t}{h D_y} \arctan \left( \frac{D_y}{\varpi h}\right), 
	\end{aligned}
\end{equation}
where $\varpi \in \left\lbrace 1, 2 \right\rbrace$, $\arctan(\cdot)$ represents the arctangent function \cite{GIS}.

\begin{IEEEproof} 
Based on (\ref{formula_2}) and (\ref{formula_10}), one has 
\begin{equation}\label{formula_15}
	\begin{aligned}
		\bar{E}_{\mathrm{LM}, \varpi} = \alpha \beta \eta P_t \mathbb{E} \left[ \frac{1}{L_{\varpi}^*} \right] = \alpha \beta \eta P_t \int_{h^2}^{h^2 + \left(\frac{D_y}{\varpi}\right)^2} \frac{1}{l}f_{L_{\varpi}^*} (l) \mathrm{d}l.  
	\end{aligned}
\end{equation}
Subsequently, substituting (\ref{formula_10}) into (\ref{formula_15}) and evaluating the definite integral yields the result in (\ref{formula_14}).
\end{IEEEproof}

\textbf{Lemma 4}. For the DDS, the average harvested energy at the UE under the LM can be derived as
\begin{equation}\label{formula_16}
	\begin{aligned}
		\bar{E}_{\mathrm{LM},3} = \alpha \beta \eta P_t \left[ \frac{2}{\Lambda h} \arctan\left( \frac{\Lambda}{h} \right) - \frac{1}{\Lambda^2} \ln\left( 1 + \frac{\Lambda^2}{h^2} \right)\right]. 
	\end{aligned}
\end{equation}

\begin{IEEEproof} 
Under the DDS configuration, and similar to the derivation in (\ref{formula_15}), we have 
\begin{equation}\label{formula_17}
	\begin{aligned}
		\bar{E}_{\mathrm{LM},3} = \alpha \beta \eta P_t \int_{h^2}^{h^2 + \Lambda^2} \frac{1}{l} \left( \frac{1}{\Lambda \sqrt{l - h^2}} - \frac{1}{\Lambda^2} \right) \mathrm{d}l. 
	\end{aligned}
\end{equation}
Then, by evaluating the integral in (\ref{formula_15}), the closed-form expression in (\ref{formula_16}) can be obtained.
\end{IEEEproof}

For the NLM, due to $\Phi \left(  P_\mathrm{in} \right)$ is a non-decreasing function of $ P_\mathrm{in} $, the harvested energy can be enhanced by appropriately optimizing $P_\mathrm{in}$. However, due to the complexity of the NLM in (\ref{formula_3}), it is challenging to derive an exact closed-form expression for the average harvested energy in (\ref{formula_4}).

To this end, an upper bound on the average harvested energy at the UE under the NLM is obtained by applying Jensen’s inequality, yielding
\begin{equation}\label{formula_18}
	\begin{aligned}
		\bar{E}_{\mathrm{NLM}} = \mathbb{E} \left[ \alpha \Phi \left( P_\mathrm{in} \right) \right] \leq \alpha \Phi \left( \mathbb{E} \left[ P_\mathrm{in} \right] \right) = \hat{E}_{\mathrm{NLM}}.
	\end{aligned}
\end{equation}

\textbf{Lemma 5}. Based on the proposed three waveguide deployment schemes and the transmission protocol, an upper bound on the average harvested energy at the UE under the NLM can be expressed as
\begin{equation}\label{formula_19}
	\begin{aligned}
		\hat{E}_{\mathrm{NLM},\nu} = \alpha \Phi \left( P_{\mathrm{in}, \nu} \right), \quad \nu \in \left\lbrace 1, 2, 3 \right\rbrace, 
	\end{aligned}
\end{equation}
where $P_{\mathrm{in}, \varpi} = \frac{ \beta \varpi P_t}{h D_y} \arctan \left( \frac{D_y}{\varpi h}\right)$, $\varpi \in \left\lbrace 1, 2 \right\rbrace$, $P_{\mathrm{in}, 3} = \beta P_t \left[ \frac{2}{\Lambda h} \arctan\left( \frac{\Lambda}{h} \right) - \frac{1}{\Lambda^2} \ln\left( 1 + \frac{\Lambda^2}{h^2} \right)\right]$.

\begin{IEEEproof} 
The proof is similar to \textbf{Lemma 3} and \textbf{Lemma 4}, which is omitted for simplicity.
\end{IEEEproof}

Next, we present the exact closed-form expressions for the average achievable rate of the UE under the three waveguide deployment schemes in the following two lemmas.

\textbf{Lemma 6}. For the EDS and CDS, the average achievable rate of the UE can be given by 
\begin{equation}\label{formula_20}
	\begin{aligned}
		\bar{R}_{\varpi} &= \frac{(1 - \alpha \beta) \varpi}{D_y \ln 2} \left[ 2 \sqrt{\mu \bar{\gamma} + h^2} \tan^{-1} \left( \frac{D_y}{\varpi \sqrt{\mu \bar{\gamma} + h^2}} \right) \right. \\ 
		& \left. \quad - 2 h \tan^{-1} \left( \frac{D_y}{\varpi h} \right) + \frac{D_y}{\varpi} \ln\left( 1 + \frac{\mu \bar{\gamma}}{h^2 + \left( \frac{D_y}{\varpi} \right)^2} \right) \right].
	\end{aligned}
\end{equation}

\begin{IEEEproof} 
Please refer to Appendix \ref{prooftwo}.
\end{IEEEproof}

\textbf{Lemma 7}. For the DDS, the average achievable rate of the UE can be derived as
\begin{equation}\label{formula_21}
	\begin{aligned}
		\bar{R}_3 = \frac{1 - \alpha \beta}{\ln 2} \left( \frac{1}{\Lambda} I_1 - \frac{1}{\Lambda^2} I_2 \right),
	\end{aligned}
\end{equation}
where 
\begin{equation}\label{formula_22}
	\begin{aligned}
		I_1 =\;& 4\sqrt{\mu \bar{\gamma} + h^2} \tan^{-1}\left( \frac{\Lambda}{\sqrt{\mu \bar{\gamma} + h^2}} \right) \\
		& \quad \ - 4h \tan^{-1}\left(\frac{\Lambda}{h}\right) + 2\Lambda \ln\left( \frac{\Lambda^2 + \mu \bar{\gamma} + h^2}{\Lambda^2 + h^2} \right),
	\end{aligned}
\end{equation}
\begin{equation}\label{formula_23}
	\begin{aligned}
		I_2 &= (h^2 + \Lambda^2) \ln\left(1 + \frac{\mu \bar{\gamma}}{h^2 + \Lambda^2} \right) + \mu \bar{\gamma} \ln(h^2 + \Lambda^2 + \mu \bar{\gamma}) \\
		&\qquad \qquad \qquad \qquad - h^2 \ln\left(1 + \frac{\mu \bar{\gamma}}{h^2} \right) - \mu \bar{\gamma} \ln(h^2 + \mu \bar{\gamma}).
	\end{aligned}
\end{equation}

\begin{IEEEproof} 
Based on (\ref{formula_6}) and (\ref{formula_10}), the average achievable rate of the UE under the DDS can be expressed as 
\begin{equation}\label{formula_24}
	\begin{aligned}
		\bar{R}_{3} &= (1 - \alpha \beta) \mathbb{E} \left[ \log_2 \left(1 + \frac{\mu \bar{\gamma}}{L_{3}^* } \right) \right] \\
		&= \frac{(1 - \alpha \beta)}{\ln 2} \left[ \frac{1}{\Lambda} \underbrace{ \int_{h^2}^{h^2 + \Lambda^2} \frac{1}{\sqrt{l - h^2}} \ln \left(1 + \frac{\mu \bar{\gamma}}{l} \right) \mathrm{d} l}_{I_1} \right. \\
		& \left. \qquad \qquad \quad \qquad - \frac{1}{\Lambda^2} \underbrace{ \int_{h^2}^{h^2 + \Lambda^2} \ln \left(1 + \frac{\mu \bar{\gamma}}{l} \right) \mathrm{d} l }_{I_2} \right] .
	\end{aligned}
\end{equation}
where $I_1$ follows from a derivation similar to Appendix \ref{prooftwo}, while $I_2$ is obtained by decomposing $\ln \left(1 + \frac{\mu \bar{\gamma}}{l} \right)$ into $\ln (l + \mu \bar{\gamma}) - \ln (l)$ and integrating each term, yielding (\ref{formula_23}). By substituting (\ref{formula_22}) and (\ref{formula_23}) into equation (\ref{formula_21}), a closed-form expression for the average achievable rate under the DDS configuration is obtained.
\end{IEEEproof}

\section{Numerical Results}
This section presents Monte-Carlo simulations to validate the accuracy of the derived analytical results. Unless otherwise specified, the simulation parameters are set as follows: $\sigma^2 = -90 $ dBm, $f_c = 28$ GHz, $D_x = 15$ m, $D_y = 10$ m, $h = 3$m, $\alpha = 0.8$, $\beta = 0.8$, $\eta = 1$, $\varphi = 20$ mW, $a = 100$ $/ \mu$W, $b = 2.9$ $\mu$W . In addition, we perform $10^6$ Monte-Carlo simulations in this paper.
\begin{figure}[!t]
	\centering
	\includegraphics[width=2.78 in]{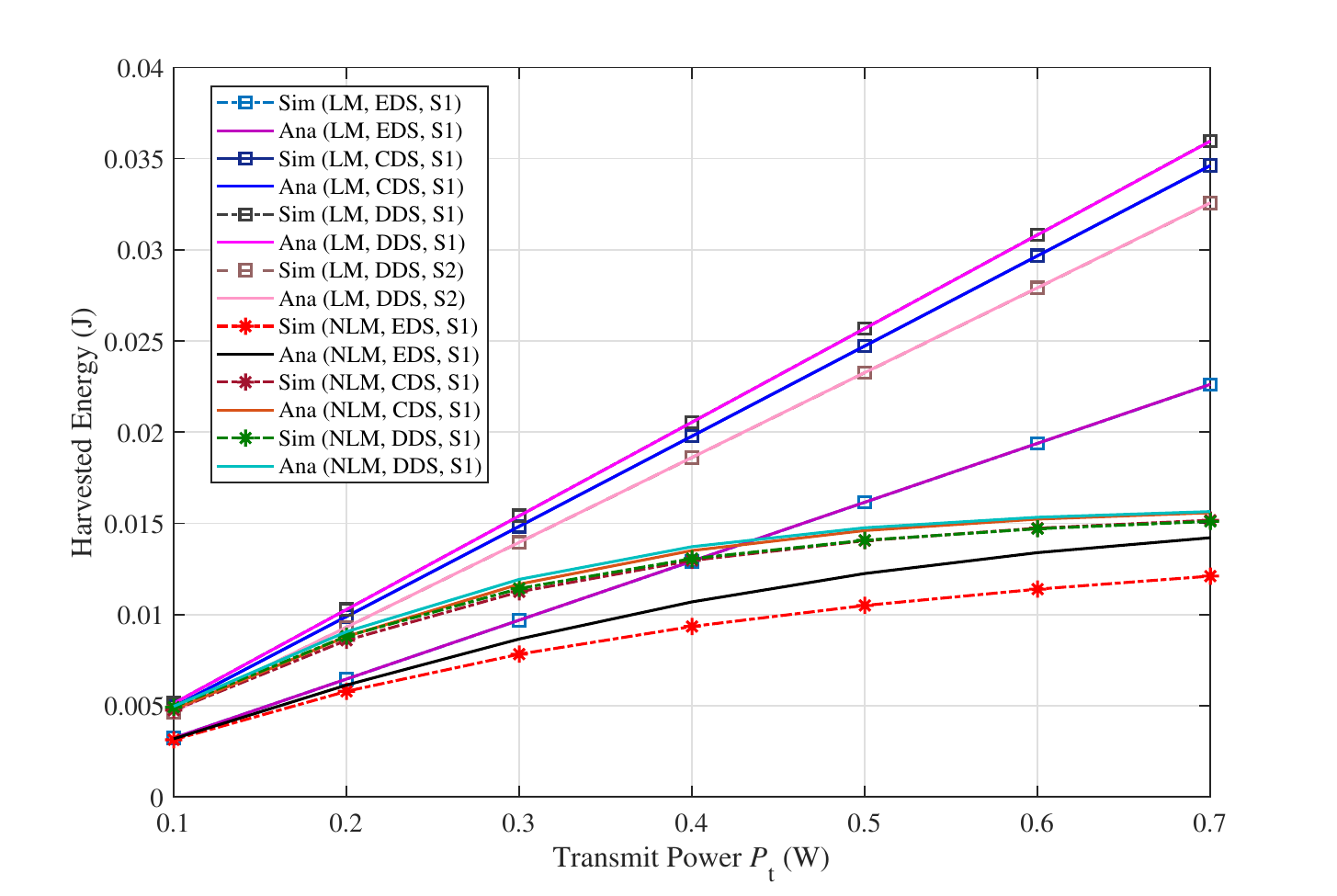}
	\caption{The average harvested energy versus the transmit power.}
	\label{Fig2}
\end{figure}
\begin{figure}[!t]
	\centering
	\includegraphics[width=2.38 in]{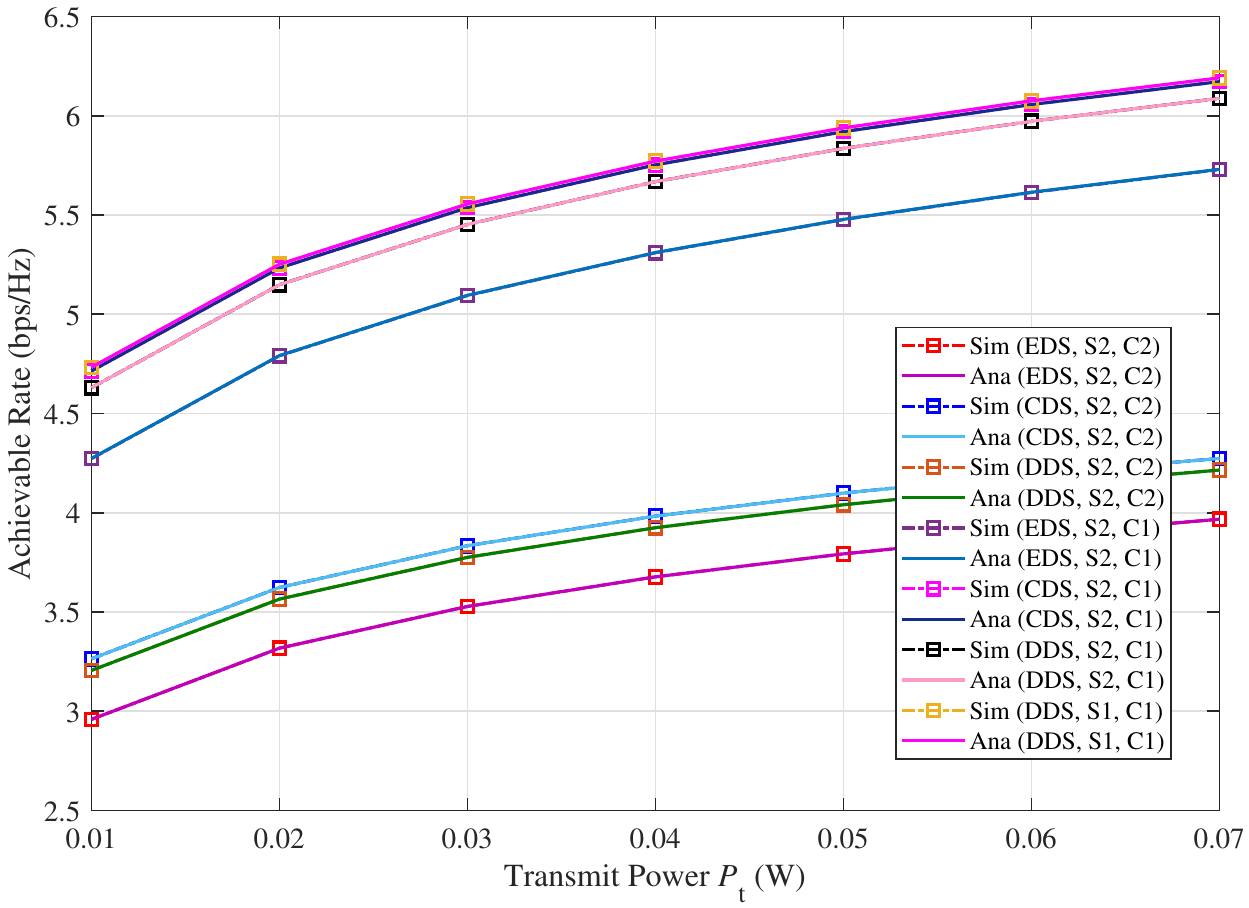}
	\caption{The average achievable rate versus the transmit power.}
	\label{Fig3}
\end{figure}
\begin{figure}[!t]
	\centering
	\includegraphics[width=2.38 in]{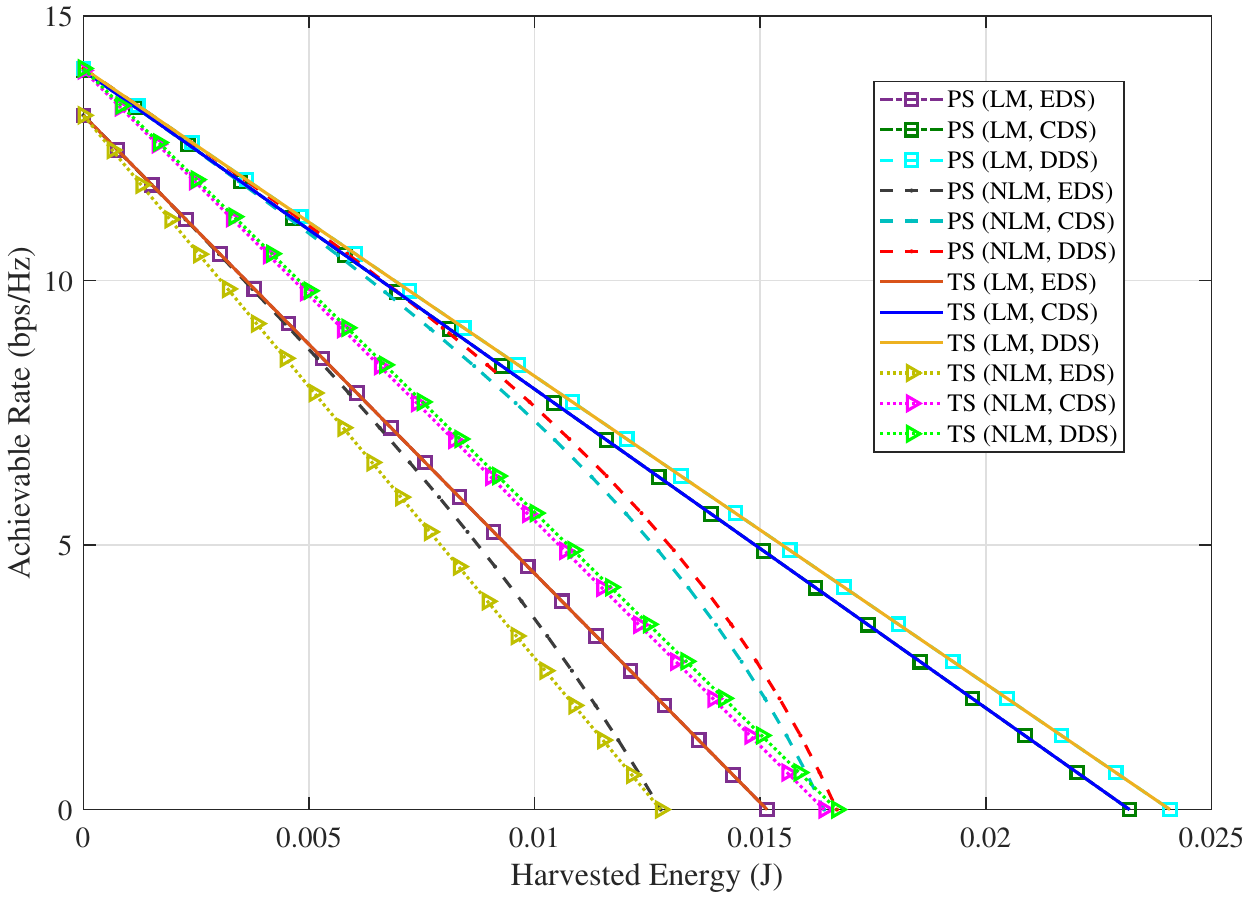}
	\caption{Energy-rate trade-off under three waveguide deployment schemes.}
	\label{Fig4}
\end{figure}

Fig. \ref{Fig2} depicts the average harvested energy under three waveguide deployment schemes and two energy harvesting models. In addition, S1 adopts a square ground area with $D_x = 8$ m and $D_y = 8$ m, while S2 uses a rectangular ground area with $D_x = 15$ m and $D_y = 8$ m. As expected, the harvested energy increases linearly with the transmit power under the LM, whereas it gradually saturates under the NLM.
Moreover, it can be observed that the energy harvested by the UE under the CDS is significantly lower than that under the EDS and DDS. However, we can observe that DDS outperforms CDS under the S1 configuration, whereas CDS achieves better performance than DDS under the S2 configuration. Therefore, selecting the optimal deployment scheme should be scenario-dependent, which will be further investigated in our future work. 

Fig. \ref{Fig3} illustrates the average achievable rate versus the transmit power under the three waveguide deployment schemes. C1 represents the case with $\alpha = 0.8$ and $\beta = 0.8$, while C2 corresponds to $\alpha = 0.6$ and $\beta = 0.6$. It can be observed that the average achievable rate increases with the transmit power. Moreover, the performance under the C2 configuration is superior to that of C1, as more power is allocated to information decoding. Similar to the observation in Fig. \ref{Fig4}, the CDS provides the lowest average achievable rate among the three deployment schemes. Moreover, while DDS outperforms CDS under the S1 configuration, its performance becomes inferior to CDS under the S2 configuration.

Fig. \ref{Fig4} shows the energy-rate trade-off under three waveguide deployment schemes, considering both linear and non-linear energy harvesting models with TS and PS protocols. The parameters are set as follows: $P_t = 0.3$ W, $D_x = 8$ m and $D_y = 8$ m. Under the LM, the energy–rate trade-off curves are identical and exhibit linear characteristics under both TS and PS protocols. In contrast, under the NLM, the trade-off remains linear for the TS protocol but becomes non-linear for the PS protocol. This is because, compared to the TS protocol, the PS protocol enables simultaneous energy harvesting and information decoding, thereby allowing more efficient utilization of available resources. Moreover, in a square room, the energy–rate region under the DDS outperforms those of the other two deployment schemes.

\section{Conclusion}
This paper investigated the performance of a novel pinching-antenna-enabled SWIPT system under three practical waveguide deployment schemes: EDS, CDS and DDS. Furthermore, a hybrid TS-PS protocol was employed to flexibly adjust the trade-off between energy harvesting and information decoding. Closed-form expressions for the average harvested energy and average achievable rate of the UE were derived under each scheme, based on the optimal positioning of the pinching-antenna. Numerical results validated the accuracy of the theoretical analysis and revealed insightful trends.

\appendices

\setcounter{equation}{0}
\renewcommand\theequation{A.\arabic{equation}}
\section{Proof of Lemma 2}\label{proofone}
\begin{figure}[!t]
	\centering
	\includegraphics[width=2.18 in]{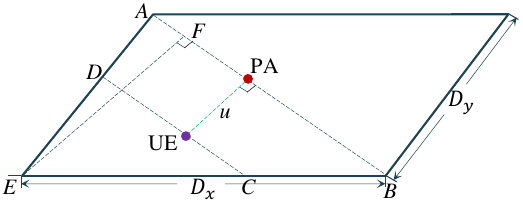}
	\caption{The ground projection of the pinching antenna.}
	\label{Fig5}
\end{figure}
For the DDS, it is quite difficult to derive the CDF of $L_3^*$ directly from its definition. To this end, we develop a clever approach for the derivation. As shown in Fig. \ref{Fig5}, we first derive the CDF of the distance $u$ between the UE and the ground projection of the pinching-antenna. Therefore, we have 
\begin{equation}\label{A1}
	\begin{aligned}
		F_u (x) = \Pr \left\lbrace u  \leq x \right\rbrace = \frac{S_{ABCD}}{S_{ABE}} = 1 - \frac{S_{CDE}}{S_{ABE}},
	\end{aligned}
\end{equation}
where $S_{\Delta}$ denotes the area of region $\Delta$. Through analytical derivation, we obtain $\frac{S_{CDE}}{S_{ABE}} = \frac{ \Lambda^2 - 2 \Lambda x + x^2 }{\Lambda^2}$. Then, substituting into (A.1) and simplifying yields $F_u (x) = \frac{2 \Lambda x - x^2}{\Lambda^2}$. Subsequently, based on (\ref{A1}) and $F_{L_3^*} (l) = \Pr \{ L_3^* \leq l \} = \Pr \{ u^2 + h^2 \leq l \} = \Pr \{ u \leq \sqrt{l - h^2} \} = F_u (\sqrt{l - h^2})$, (\ref{formula_12}) can be obtained. Finally, $f_{L_3^*} (l)$ can be obtained by taking the first derivative of (\ref{formula_12}). The proof is completed.

\setcounter{equation}{0}
\renewcommand\theequation{B.\arabic{equation}}
\section{Proof of Lemma 6}\label{prooftwo}

Based on (\ref{formula_6}) and (\ref{formula_10}), the average achievable rate of the UE under the EDS and CDS can be given by 
\begin{equation}\label{B1}
	\begin{aligned}
		\bar{R}_{\varpi} &= (1 - \alpha \beta) \mathbb{E} \left[ \log_2 \left(1 + \frac{\mu \bar{\gamma}}{L_{\varpi}^* } \right) \right] \\
		&= \frac{(1 - \alpha \beta) \varpi}{2 D_y \ln 2} \int_{h^2}^{h^2 + \left(\frac{D_y}{\varpi}\right)^2} \ln \left(1 + \frac{\mu \bar{\gamma}}{l} \right) \frac{1}{ \sqrt{l - h^2}}\mathrm{d} l \\
		&{\overset{\left(l = h^2 + t^2 \right)}{=}} \frac{(1 - \alpha \beta) \varpi}{D_y \ln 2} \underbrace{ \int_{0}^{\frac{D_y}{\varpi}} \ln \left( 1 + \frac{\mu \bar{\gamma}}{h^2 + t^2}\right)}_I \mathrm{d} t.
	\end{aligned}
\end{equation}
where $\varpi \in \left\lbrace 1, 2\right\rbrace $. Then, by applying integration by parts to $I$, we obtain
\begin{equation}\label{B2}
	\begin{aligned}
		I &= 2 \sqrt{\mu \bar{\gamma} + h^2} \tan^{-1} \left( \frac{D_y}{\varpi \sqrt{\mu \bar{\gamma} + h^2}} \right) - 2 h \tan^{-1} \left( \frac{D_y}{\varpi h} \right) \\ 
		&\qquad \qquad \qquad \qquad \qquad + \frac{D_y}{\varpi} \ln\left( 1 + \frac{\mu \bar{\gamma}}{h^2 + \left( \frac{D_y}{\varpi} \right)^2} \right).
	\end{aligned}
\end{equation}

Finally, by substituting (\ref{B2}) into equation (\ref{B1}), (\ref{formula_20}) is obtained. Here, the proof is completed.

\bibliographystyle{IEEEtran}
\bibliography{MyRef}

\begin{thebibliography}{10}
\providecommand{\url}[1]{#1}
\csname url@samestyle\endcsname
\providecommand{\newblock}{\relax}
\providecommand{\bibinfo}[2]{#2}
\providecommand{\BIBentrySTDinterwordspacing}{\spaceskip=0pt\relax}
\providecommand{\BIBentryALTinterwordstretchfactor}{4}
\providecommand{\BIBentryALTinterwordspacing}{\spaceskip=\fontdimen2\font plus
\BIBentryALTinterwordstretchfactor\fontdimen3\font minus
  \fontdimen4\font\relax}
\providecommand{\BIBforeignlanguage}[2]{{%
\expandafter\ifx\csname l@#1\endcsname\relax
\typeout{** WARNING: IEEEtran.bst: No hyphenation pattern has been}%
\typeout{** loaded for the language `#1'. Using the pattern for}%
\typeout{** the default language instead.}%
\else
\language=\csname l@#1\endcsname
\fi
#2}}
\providecommand{\BIBdecl}{\relax}
\BIBdecl

\bibitem{HJYK}
J.~Hu, K.~Yang, G.~Wen, and L.~Hanzo, ``Integrated data and energy
  communication network: A comprehensive survey,'' \emph{IEEE Communications
  Surveys \& Tutorials}, vol.~20, no.~4, pp. 3169--3219, 2018.

\bibitem{WQZR}
Q.~Wu and R.~Zhang, ``Intelligent reflecting surface enhanced wireless network
  via joint active and passive beamforming,'' \emph{IEEE Transactions on
  Wireless Communications}, vol.~18, no.~11, pp. 5394--5409, 2019.

\bibitem{WKK}
K.-K. Wong, A.~Shojaeifard, K.-F. Tong, and Y.~Zhang, ``Fluid antenna
  systems,'' \emph{IEEE Transactions on Wireless Communications}, vol.~20,
  no.~3, pp. 1950--1962, 2021.

\bibitem{ZLMW}
L.~Zhu, W.~Ma, and R.~Zhang, ``Modeling and performance analysis for movable
  antenna enabled wireless communications,'' \emph{IEEE Transactions on
  Wireless Communications}, vol.~23, no.~6, pp. 6234--6250, 2024.

\bibitem{AFHY}
A.~Fukuda, H.~Yamamoto, H.~Okazaki, Y.~Suzuki, and K.~Kawai, ``Pinching
  antenna-using a dielectric waveguide as an antenna,'' \emph{NTT DOCOMO Tech.
  J.}, vol.~23, no.~3, pp. 5--12, 2022.

\bibitem{LYWZ}
Y.~Liu, Z.~Wang, X.~Mu, C.~Ouyang, X.~Xu, and Z.~Ding, ``Pinching-antenna
  systems (pass): Architecture designs, opportunities, and outlook,'' 2025,
  [Online]. Available: \url{https://doi.org/10.48550/arXiv.2501.18409}.

\bibitem{DZSR}
Z.~Ding, R.~Schober, and H.~Vincent~Poor, ``Flexible-antenna systems: A
  pinching-antenna perspective,'' \emph{IEEE Transactions on Communications},
  pp. 1--1, 2025.

\bibitem{XJWJ}
J.~Xiao, J.~Wang, and Y.~Liu, ``Channel estimation for pinching-antenna systems
  (pass),'' \emph{IEEE Communications Letters}, pp. 1--1, 2025.

\bibitem{ZZYZ}
Z.~Zhou, Z.~Yang, G.~Chen, and Z.~Ding, ``Sum-rate maximization for
  noma-assisted pinching-antenna systems,'' \emph{IEEE Wireless Communications
  Letters}, pp. 1--1, 2025.

\bibitem{TDTSA}
D.~Tyrovolas, S.~A. Tegos, P.~D. Diamantoulakis, S.~Ioannidis, C.~K. Liaskos,
  and G.~K. Karagiannidis, ``Performance analysis of pinching-antenna
  systems,'' \emph{IEEE Transactions on Cognitive Communications and
  Networking}, pp. 1--1, 2025.

\bibitem{GIS}
I.~S. Gradshteyn and I.~M. Ryzhik, \emph{Table of Integrals, Series, and
  Products, \rm{7th}.}\hskip 1em plus 0.5em minus 0.4em\relax San Diego, CA:
  Academic Press, 2007.

\end{thebibliography}

\end{document}